\documentclass[%
 aip,
 sd,%
 amsmath,amssymb,
 reprint,%
]{revtex4-1}

\usepackage{graphicx}% Include figure files
\usepackage{dcolumn}% Align table columns on decimal point
\usepackage{bm}% bold math

\begin{document}

\preprint{AIP/123-QED}

\title[]{Path-Integral Isomorphic Hamiltonian for Including Nuclear Quantum Effects in Non-adiabatic Dynamics}% Force line breaks with \\
%{Classical Isomorphic Hamiltonian:\\ Including Nuclear Quantum Effects in Non-adiabatic Dynamics}% Force line breaks with \\
%\thanks{Footnote to title of article.}

\author{Xuecheng Tao}
\author{Philip Shushkov}%
%\email{Second.Author@institution.edu.}

\author{Thomas F. Miller III}
\thanks{Electronic mail: tfm@caltech.edu.}
%\homepage{http://www.Second.institution.edu/~Charlie.Author.}
\affiliation{
Division of Chemistry and Chemical Engineering, California Institute of Technology,\\ Pasadena,
California 91125, USA %\\This line break forced with \textbackslash\textbackslash
}%

\date{\today}% It is always \today, today,

\begin{abstract}
We describe a path-integral approach for including nuclear quantum effects in non-adiabatic chemical dynamics simulations. For a general physical system with multiple electronic energy levels, a corresponding isomorphic Hamiltonian is introduced, such that Boltzmann sampling of the isomorphic Hamiltonian with classical nuclear degrees of freedom yields the exact quantum Boltzmann distribution for the original physical system.  In the limit of a single electronic energy level, the isomorphic Hamiltonian reduces to the familiar cases of either ring polymer molecular dynamics (RPMD) or centroid molecular dynamics Hamiltonians, depending on implementation.  An advantage of the isomorphic Hamiltonian is that it can easily be combined with existing mixed quantum-classical dynamics methods, such as surface hopping or Ehrenfest dynamics, to enable the simulation of electronically non-adiabatic processes with nuclear quantum effects. We present numerical applications of the isomorphic Hamiltonian to model two- and three-level systems, with encouraging results that include improvement upon a previously reported combination of RPMD with surface hopping in the deep-tunneling regime.

\end{abstract}

% Valid PACS numbers may be entered using the \verb+\pacs{#1}+ command.
\pacs{Valid PACS appear here}% PACS, the Physics and Astronomy
%Use showkeys class option if keyword
\maketitle

%\begin{quotation}
%The ``lead paragraph'' is encapsulated with the \LaTeX\ 
%\end{quotation}

\section{\label{sec:level1}Introduction}

Chemical processes that involve transitions among different electronic states play a central role in photo-induced,\cite{xu2014photodissociation, levine2007isomerization} redox,\cite{shih2008tryptophan, fujita2004mimicking} and collisional processes.\cite{nahler2008inverse, yao2017dynamic} 
Widely used mixed quantum-classical (MQC) methods - including Ehrenfest dynamics\cite{ehrenfest1933phase} and surface hopping\cite{tully1990molecular} - have been developed for the simulation of  electronically non-adiabatic processes in cases for which the nuclei can be described using classical mechanics.  However, nuclear quantum effects are important in many electronically non-adiabatic processes,\cite{xie2016nonadiabatic, hammes2010theory, huynh2007proton, migliore2014biochemistry} creating the need for new methods that robustly and accurately describe the interplay between nuclear and electronic quantum mechanical effects.

For chemical dynamics on a single electronic surface, 
approximate methods based on imaginary-time Feynman path integrals \cite{feynman1965quantum, chandler1981exploiting} 
 have proven useful for describing nuclear quantization. 
These methods include ring-polymer molecular dynamics\cite{craig2004quantum, habershon2013ring}
 (RPMD) and centroid molecular dynamics\cite{cao1994formulation2, voth1996path,  jang1999derivation}
  (CMD), which involve classical molecular dynamics trajectories governed by an isomorphic Hamiltonian that includes the effects of zero-point energy and tunneling.  RPMD and CMD  exhibit various exact formal properties, including time-reversibility and  preservation of the quantum Boltzmann distribution for the physical system, and RPMD additionally recovers semiclassical instanton rate theory in the deep-tunneling regime.  \cite{richardson2009ring} 
The simplicity and robustness of these path-integral-based methods has led to the development of mature technologies \cite{rossi2014remove, ceriotti2014pi, markland2008efficient, marsalek2016ab} and enables the study of complex systems.\cite{miller2005quantum, miller2005quantumwater,  habershon2008comparison, boekelheide2011dynamics}

These successes 
motivate the development of path-integral-based methods for describing electronically non-adiabatic dynamics. 
Previous work includes non-adiabatic extensions of instanton theory, \cite{cao1995computation, schwieters1998semiclassical, jang2001nonadiabatic} CMD,\cite{schwieters1999extension, liao2002centroid} and RPMD. \cite{ananth2013mapping, richardson2013communication, shushkov2012ring, menzeleev2014kinetically, kretchmer2017kinetically, kretchmer2015tipping, kretchmer2013direct}
A unifying feature of these previous efforts 
is that they employ a case-specific development strategy, in which path-integral  quantization of the nuclei is specifically tailored for combination with a particular approximation to the electronically non-adiabatic dynamics, such as instanton theory,\cite{cao1995computation, schwieters1998semiclassical, jang2001nonadiabatic, menzeleev2014kinetically, kretchmer2017kinetically}
surface-hopping,\cite{shushkov2012ring, shakib2017ring} linearized semiclassical,\cite{ananth2010exact,ananth2007semiclassical, sun1998semiclassical, sun1997semiclassical, huo2010iterative, huo2013communication} 
or other approximation. 
This strategy typically limits each resulting method to the application domain for which the associated non-adiabatic dynamics approximation is valid. 

The current work employs an alternative strategy to take full advantage of the diversity of previously developed MQC methods for describing non-adiabatic dynamics.
We use path integration to obtain a general isomorphic Hamiltonian that incorporates nuclear quantization and that can be easily combined with any MQC method.  As will be shown, this leads to a variety of promising, new dynamics methods that retain the simplicity and robustness of both imaginary-time path-integrals for nuclear quantization and the parent MQC method.  
In the following, we derive the new isomorphic Hamiltonian, and we present applications of it in combination with non-adiabatic dynamics based on either surface hopping\cite{tully1990molecular} or the quantum-classical Liouville equation.\cite{kapral1999mixed, nielsen2000mixed}  These results illustrate the flexibility with which the  isomorphic Hamiltonian may be employed, as well as implementations that are readily applicable for the study of complex systems.

\section{THEORY}
\label{TheorySection}

We begin by reviewing the path-integral-based RPMD and CMD methods, which employ an isomorphic Hamiltonian for the description of quantized nuclear dynamics in electronically adiabatic systems.   We then extend this approach to obtain an isomorphic Hamiltonian for the description of quantized nuclear dynamics  involving multiple electronic states.

\subsection{Isomorphic Hamiltonian for one-level systems: RPMD and CMD}
For a system obeying the Born-Oppenheimer approximation in the electronic ground state, we consider the Hamiltonian operator
\begin{equation} \label{system}
\hat{H}=\frac{p^2}{2m}+V(x),
\end{equation}
where $x$, $p$, and $m$ are the nuclear position, momentum, and mass, respectively, and  $V(x)$ is the potential energy surface.  Throughout this work, results will be presented for a single nuclear degree of freedom; generalization to multiple dimensions is straightforward.  

The path-integral discretization of the quantum mechanical canonical partition function for this system is given by \cite{feynman1965quantum, chandler1981exploiting, parrinello1984study} 
\begin{eqnarray}
\label{pipf}
Q&&=\text{tr}[e^{-\beta  \hat{H}}] \nonumber  \\
&&=\lim_{n\rightarrow\infty} \left(\frac{n}{2\pi \hbar}  \right)^{n}  \!\! \int \!d\textbf{x} \int \! d\textbf{p}\ e^{-\beta H_n^{\rm iso}(\textbf{x}, \textbf{p})},
\end{eqnarray}
where $\beta$ is the reciprocal temperature,  $n$ is the number of ring-polymer beads in the path-integral discretization, $\textbf{x}=\{x^{(1)}, x^{(2)}, \ldots, x^{(n)}\}$ is the vector of ring-polymer positions such that $x^{(1)}=x^{(n+1)}$, and $\textbf{p}$ is the vector of ring-polymer momenta. $H_n^{\rm iso}$ is the ring-polymer Hamiltonian (see Appendix A)
\begin{equation} \label{rpmdH}
H_n^{\rm iso}(\textbf{x}, \textbf{p}) = \sum_{\alpha=1}^n \frac{p_{\alpha}^2}{2m_n}  + U_{\text{spr}}(\textbf{x}) + \frac1n \sum_{\alpha=1}^n V(x_{\alpha}), 
\end{equation}
which includes the inter-bead potential
\begin{equation}
\label{Uspr}
U_{\text{spr}}(\textbf{x}) =\frac{1}{2} \, m_n \, \omega_n^2 \sum_{\alpha=1}^n   (x_{\alpha} - x_{(\alpha+1)})^2,
\end{equation}
where 
$m_n=m/n$, $\omega_n=(\beta_n\hbar)^{-1}$, and  $\beta_n=\beta/n$.

Approximate real-time quantum dynamics is obtained in the RPMD method\cite{craig2004quantum} by running classical molecular dynamics trajectories associated with the ring-polymer Hamiltonian, which are given by
\begin{eqnarray}
\label{rpmd_eom}
\dot{x}_{\alpha}&=&p_{\alpha}/m_n\\
\dot{p}_{\alpha}&=&m_n \omega_n^2\left(x_{(\alpha+1)}+x_{(\alpha-1)}-2x_{\alpha}\right)
- \frac1n \frac{\partial}{\partial x_{\alpha}} V\!\left(x_{\alpha}\right) \nonumber
\end{eqnarray}
for $\alpha=1,\ldots,n$.  

Equation \ref{pipf} can be further reduced with respect to the intra-ring-polymer degrees of freedom, yielding 
\begin{equation}
Q= \left(\frac{n}{2\pi \hbar}  \right)^{n} \!\! \int \!d\bar{x} \int \! d\bar{p}\ e^{-\beta \bar{H}^{\rm iso}(\bar{x}, \bar{p})},
\end{equation}
where $\bar{H}^{\rm iso}$ is the centroid Hamiltonian
\begin{equation}  \label{cmdH}
\bar{H}^{\rm iso}(\bar{x}, \bar{p}) =  \frac{\bar{p}^2}{2m}  + \bar{V}(\bar{x})
\end{equation}
which  includes the centroid potential of mean force
\begin{equation} 
e^{-\beta \bar{V}(\bar{x})} \! \propto \! \lim_{n\rightarrow\infty} \! \int \!\!\!d\textbf{x}\!\! \int \!\!\! d\textbf{p}\ \delta (\bar{x}-\frac1n \!\sum_{\alpha} x_{\alpha}) e^{-\beta H_n^{\rm iso}(\textbf{x}, \textbf{p})}.
\end{equation}
Approximate real-time quantum dynamics is obtained in the CMD method\cite{jang1999derivation} by running classical molecular dynamics trajectories associated with the centroid Hamiltonian, which are given by
\begin{eqnarray}
\label{cmd_eom}
\dot{\bar{x}}&=&\bar{p}/m\\
\dot{\bar{p}}&=&-\frac{\partial}{\partial \bar{x}} \bar{V}\!\left(\bar{x}\right). \nonumber
\end{eqnarray}

Both Eqs.~\ref{rpmdH} and \ref{cmdH} provide an isomorphic Hamiltonian for the one-level physical system described by  Eq.~\ref{system}, in the sense that \textit{classical mechanical}  trajectories associated with the isomorphic Hamiltonian yield the approximate \textit{quantum mechanical} time-evolution for the physical system.  
 Moreover, classical Boltzmann sampling of the isomorphic Hamiltonian (i.e., by running the classical trajectories in  Eqs.~\ref{rpmd_eom} or \ref{cmd_eom} in contact with a thermal bath) 
rigorously preserves the exact quantum Boltzmann statistics associated with the physical system.  In the following, we  derive both RPMD and CMD versions of the corresponding isomorphic Hamiltonian 
for physical systems involving multiple electronic surfaces, with the RPMD version presented in the main text and the CMD version in Appendix B.

\subsection{Isomorphic Hamiltonian for multi-level systems}

\subsubsection{Path-integral discretization}

Consider the Hamiltonian in the diabatic representation for a system with $f$ electronic energy levels,
\begin{eqnarray} \label{systemnon}
\hat{H}&=&\frac{p^2}{2m}+\hat{V}(x)\\
&=&\frac{p^2}{2m}+\left[\!\!\!\!
\begin{matrix}
&V_{1}(x)    &K_{12}(x) &\cdots  &K_{1f}(x) \\
&K_{12}(x)      &V_2(x)   &\cdots  &K_{2f}(x) \\
&\vdots &\vdots &\ddots &\vdots \\
&K_{1f}(x)   &K_{2f}(x)  &\cdots &V_f(x)   \\
\end{matrix}
\right].\nonumber
\end{eqnarray}
Discretizing the partition function with respect to both electronic state, $i$, and nuclear position, $x$, and employing a Trotter factorization such as
\begin{eqnarray}
e^{-\beta_n \hat{H}} &=& e^{-\beta_n \hat{V}/2 } e^{-\beta_n \hat{T} } e^{-\beta_n \hat{V}/2 } + \mathcal{O}(\beta_n^3), 
\end{eqnarray}
we obtain the path-integral representation 
\begin{eqnarray} \label{isonon}
Q&=&\lim_{n\rightarrow\infty}  \left(\frac{n}{2\pi \hbar}  \right)^{n}  \\
 && \times \!\! \int \!d\textbf{x} \int \! d\textbf{p}\ 
e^{-\beta ( \sum_{\alpha=1}^n \frac{p_{\alpha}^2}{2m_n} + U_{\text{spr}}(\textbf{x}))} \mu(\textbf{x}),\nonumber
\end{eqnarray}
where
\begin{equation} \label{mu}
\mu(\textbf{x})=\textrm{tr}_{\textrm{e}}\left[\ \prod_{\alpha=1}^{n} e^{-\beta_n \hat{V} (x_{\alpha}) } \right].
\end{equation}
The subscript `e' in Eq.~\ref{mu} indicates the trace taken  over only the electronic states. 
Although path-integral discretization of multi-level systems can also be performed in the adiabatic representation,\cite{schmidt2007path} the diabatic representation employed here is particularly convenient.

Note that $\mu$, which describes the statistical weight of a given ring-polymer nuclear configuration after thermally averaging over the electronic states, is a familiar and easily evaluated quantity.  It is the central object in the Schwieters-Voth non-adiabatic instanton theory \cite{schwieters1998semiclassical, schwieters1999extension, jang2001nonadiabatic} 
and  mean-field non-adiabatic RPMD,\cite{duke2017mean, menzeleev2014kinetically, hele2005an} 
both of which provide a thermally averaged (i.e., mean-field) description of the electronically non-adiabatic dynamics.
Moreover, as is discussed in Appendix C,
$\mu$ is  non-negative  when evaluated in the limit of large $n$,
and both $\mu$ and its derivative with respect to the ring-polymer nuclear coordinates can be evaluated using 
$\mathcal{O}(n)$ operations.

\subsubsection{The Isomorphic Hamiltonian}

We now address the central goal of this work: 
Given the physical system associated with the $f$-level Hamiltonian in Eq.~\ref{systemnon}, 
determine the corresponding $f$-level isomorphic Hamiltonian for which classical Boltzmann sampling 
of the nuclear degrees of freedom yields the exact quantum Boltzmann distribution for the physical Hamiltonian.
It follows from Eq.~\ref{isonon} that this requirement 
is satisfied by an  isomorphic Hamiltonian of the form
\begin{equation}
\label{isoHam}
\hat{H}^{\textrm{iso}}_n(\textbf{x}, \textbf{p})=
 \sum_{\alpha=1}^n \frac{p_{\alpha}^2}{2m_n} + 
U_{\text{spr}}(\textbf{x}) +
\hat{V}^{\textrm{iso}}(\textbf{x}),
\end{equation}
where   $\hat{V}^{\textrm{iso}}$ is the isomorphic potential energy  given by the $f\times f$ matrix that obeys
\begin{equation} \label{isox}
\textrm{tr}_{\textrm{e}}\left[  e^{-\beta \hat{V}^{\textrm{iso}}(\textbf{x}) } \right] \equiv   \mu(\textbf{x}). \\
\end{equation}
%In the following subsections, we specify the elements of $\hat{V}^{\textrm{iso}}$, first for the special case of a two-level system and then for a general multi-level system. {\color{blue}{[REVIEWER I POINT II, IV; We would like to emphasize here the isomorphic potential we derived still has the multi-dimension nature, namely, has the same number of the surfaces and couplings as the physical potential, different from the single thermally-averaged effective surface defined in the previous mean-field approaches. }}

\subsubsection{Special case of a two-level system}

For a system with two electronic states ($f=2$), the isomorphic potential energy  has the form
\begin{equation}
\label{isoV_2s}
\hat{V}^{\text{iso}}(\textbf{x})=
\left[
\begin{matrix}
&V_{1}^{\text{iso}}(\textbf{x})     &K_{12}^{\text{iso}}(\textbf{x}) \\
&K_{12}^{\text{iso}}(\textbf{x})       &  V_{2}^{\text{iso}}(\textbf{x})  \\
\end{matrix}
\right]. \\
\end{equation}
Given the symmetry of the off-diagonal term, the matrix has only three independent  elements at any given ring-polymer configuration.
To specify the two diagonal terms, we require that the usual RPMD surfaces be recovered in the regime of zero electronic coupling, such that
\begin{equation} \label{diagdef}
V_{i}^{\text{iso}}(\textbf{x}) = \frac{1}{n} \sum_{\alpha=1}^n V_i(x_{\alpha}).
\end{equation}
The only remaining term is  the off-diagonal isomorphic coupling, $K_{ij}^{\text{iso}}(\textbf{x})$, which must satisfy Eq.~\ref{isox}, such that
\begin{eqnarray} \label{offdiagdef}
\left(K_{ij}^{\text{iso}}(\textbf{x})\right)^2  &=&  {\rm acosh}^2 \left[ \, e^{\frac{\beta}{2} \left( V_{i}^{\text{iso}}(\textbf{x}) +V_{j}^{\text{iso}}(\textbf{x}) \right)} \, \mu_{ij}(\textbf{x}) /2   \right] / \beta^2 \nonumber\\ 
&&\quad
 - \left(  V_{i}^{\text{iso}}(\textbf{x}) -V_{j}^{\text{iso}}(\textbf{x}) \right)^2 / 4,
\end{eqnarray}
where
\begin{equation}\label{muij}
\mu_{ij}(\textbf{x})=\textrm{tr}_{\textrm{e}}\!\!\left[\prod_{\alpha=1}^{n} {\rm exp}\!\left(\!\!-\beta_n\! 
\left[\!\!
\begin{matrix}
&V_{i}(x_\alpha)     &K_{ij}(x_\alpha) \\
&K_{ij}(x_\alpha)       &  V_{j}(x_\alpha)  \\
\end{matrix}
\right]
 \right)\right].
\end{equation}
For the case of a two-level system, $\mu_{ij}(\textbf{x}) = \mu(\textbf{x})$, where the latter is defined in Eq.~\ref{mu}.
Eq.~\ref{offdiagdef} fully specifies $K_{ij}^{\text{iso}}(\textbf{x})$ to within an absolute sign, which we take  to be equal to that of the physical potential coupling evaluated at the ring-polymer centroid position,  ${\rm sgn}(K_{ij}(\bar{x}))$.

For a two-level system, the isomorphic Hamiltonian is given by Eq.~\ref{isoHam} and Eqs.~\ref{isoV_2s}-\ref{offdiagdef}.  
Inspection of the matrix elements of the isomorphic potential reveals that the diagonal matrix elements (Eq.~\ref{diagdef}) include  RPMD-like corrections to the diabatic potential energy surfaces, while the off-diagonal elements (Eq.~\ref{offdiagdef})   include the effect of nuclear quantization on the pairwise (i.e., two-body) coupling between the electronic states.  Before  discussing other properties of the isomorphic Hamiltonian, we generalize it to  multi-level systems.

%\newpage
\subsubsection{General case of a multi-level system}
\label{sec:multilevel}

Following the two-level case, we now present the generalization of 
the isomorphic Hamiltonian to systems with $f>2$.  We define an $f\times f$ potential energy matrix
\begin{equation}
\label{V2b}
\hat{V}^{\text{iso}}_{\textrm{2-body}}(\textbf{x})=
\left[
\begin{matrix}
& V_{1}^{\rm iso}(\textbf{x})    &K_{12}^{\text{iso}}(\textbf{x}) &\cdots  &K_{1f}^{\text{iso}}(\textbf{x}) \\
&K_{12}^{\text{iso}}(\textbf{x})      & V_{2}^{\text{iso}}(\textbf{x})   &\cdots  &K_{2f}^{\text{iso}}(\textbf{x}) \\
&\vdots &\vdots &  \ddots &\vdots \\
&K_{1f}^{\text{iso}}(\textbf{x})   &K_{2f}^{\text{iso}}(\textbf{x})  &\cdots & V_{f}^{\text{iso}}(\textbf{x}) \\
\end{matrix}
\right]
\end{equation}
for which the diagonal and off-diagonal terms are defined   in Eqs.~\ref{diagdef} and \ref{offdiagdef}.  And finally, to ensure that Eq.~\ref{isox} is satisfied, we define the isomorphic potential energy to be 
\begin{equation}
\label{Viso}
\hat{V}^{\text{iso}}(\textbf{x})= \hat{V}^{\text{iso}}_{\textrm{2-body}}(\textbf{x}) +  V^{\text{iso}}_{\textrm{many-body}}(\textbf{x}), 
\end{equation}
where
\begin{equation}
\label{Vmb}
V^{\text{iso}}_{\textrm{many-body}}(\textbf{x}) 
=-\frac1\beta\ \textrm{ln} \left[\frac{ \mu(\textbf{x})}
{ \textrm{tr}_{\textrm{e}}\!\left[  e^{-\beta \hat{V}^{\text{iso}}_{\textrm{2-body}}(\textbf{x}) } \right]   } \right]
\end{equation}
and $\mu(\textbf{x})$ is defined in Eq.~\ref{mu}.

Combined with Eq.~\ref{isoHam}, Eqs.~\ref{V2b}-\ref{Vmb} present the central result of this work: the  isomorphic Hamiltonian for a  general multi-level system.
We now point out a number of important properties that make the isomorphic Hamiltonian amenable to the description of complex, multi-level systems, much like standard RPMD and CMD are amenable to the description of complex, one-level systems.

First, the isomorphic Hamiltonian 
can immediately be employed with any MQC method for describing nonadiabatic dynamics; by simply running the MQC dynamics on the isomorphic Hamiltonian, nuclear quantum effects are included via the path-integral description.  Naturally, the dynamics run on the isomorphic Hamiltonian will inherit the strengths and weaknesses of the MQC method that is employed.
As is  illustrated in the Results section, the MQC dynamics can either be run directly using
the diabatic representation or by diagonalizing it to obtain the corresponding adiabatic states and derivative couplings.

Second, by construction, the   isomorphic Hamiltonian satisfies the   requirement that classical Boltzmann sampling 
of the nuclear degrees of freedom 
yields the exact quantum Boltzmann distribution for the physical system.  It employs a path-integral discretization that involves no approximation to the quantum statistics of the system.
For an (idealized) MQC method for which the equations of motion rigorously preserve the MQC Boltzmann ensemble, then  running the corresponding dynamics on the isomorphic Hamiltonian would rigorously preserve the exact quantum Boltzmann distribution; however, we note that most MQC methods do not rigorously preserve the MQC Boltzmann ensemble.\cite{schmidt2008mixed}

Third, as for standard RPMD, evaluation of the matrix elements in the isomorphic Hamiltonian is numerically robust and scales linearly in cost with the number of ring-polymer beads.  Quantities that arise in the evaluation of the isomorphic Hamiltonian, such as $\mu(\textbf{x})$, $\mu_{ij}(\textbf{x})$, or  $ \textrm{tr}_{\textrm{e}}\!\left[  e^{-\beta \hat{V}^{\text{iso}}_{\textrm{2-body}}(\textbf{x}) } \right]$ (and their derivatives with respect to nuclear position), can be obtained from simple diagonalization of an $f\times f$ matrix or with $\mathcal{O}(n)$ operations.  
Furthermore, the argument of the logartithm in Eq.~\ref{Vmb} involves a ratio of positive quantities and is thus well behaved. 
It should be noted that the numerical robustness of the isomorphic Hamiltonian is an important and non-trivial feature; whereas evaluation of the path-integral representation for the underlying density matrix of a many-level system generally  gives rise to a numerical sign problem,\cite{chandler1987introduction} we have  expressed the isomorphic Hamiltonian in terms of non-oscillatory quantities.

We further note that the isomorphic Hamiltonian obeys various satisfying limits. 
In the classical mechanical limit for the physical nuclei (i.e., the 1-bead ring polymer limit), the isomorphic Hamiltonian reduces to the original physical Hamiltonian in Eq.~\ref{systemnon}.
In the limit of zero coupling among the states in the physical system (i.e., when $K_{ij}=0$), the isomorphic Hamiltonian reduces to the standard RPMD Hamiltonian for the diabatic potential energy surfaces.
Finally, in the limit for which the electronic states only couple via separate pairs, $V^{\text{iso}}_{\textrm{many-body}}(\textbf{x}) =0$,  the many-level isomorphic Hamilton simply reduces to the previously discussed two-level result.  In this sense, $\hat{V}^{\text{iso}}_{\textrm{2-body}}(\textbf{x})$
 includes the effect of nuclear quantization on the pairwise (i.e., two-body) coupling between the electronic states, whereas $V^{\text{iso}}_{\textrm{many-body}}(\textbf{x})$
 provides a mean-field many-body coupling between the electronic states due to nuclear quantization.
 As will be seen in the results, this many-body coupling is found to be much smaller than the two-body coupling, but inclusion of the many-body term is necessary to rigorously preserve the  quantum Boltzmann statistics.

Finally, we note that the specification of the matrix elements of the isomorphic potential presented here is not unique.
 For example, direct inversion of the electronic density matrix within the trace operation of Eq.~\ref{mu} was explored and found to be numerically ill-conditioned. 
Other alternative choices that satisfy the condition in Eq.~\ref{isox} may be devised, although
 any revision should both preserve the formal properties listed above and improve upon the numerical results presented in the Results section. 
%{\color{blue}{For example, directly taking the matrix logrithm of the reduced density matrix, the object inside the trace operator in the Eq.~\ref{mu}, is a natural thought at a first glance and the quantum Boltzmann statistics is also preserved. However, the isomorphic Hamiltonian defined in this way creates unphysical well in the surfaces, originated from the fact that the contributions from the off-diagonal potential coupling appear in the diagonal elements of the isomorphic potential. In the opposite side, our specific realization confines all the coupling contributions to the off-diagonal elements of the isomorphic potential and is in consistent with our understanding of the diabatic surfaces, whose chemical character is fixed and should not dependent on any other electronic surfaces. Besides, }} 
 We do recognize that a representation-invariant specification of the matrix elements of the isomorphic potential would be a worthy goal for future development.
 Similarly, we recognize the mathematical possibility that the RHS of Eq.~\ref{offdiagdef} may become negative in our specification 
 (although we have found no such case in which this occurs), and we note that the positivity of $\mu$ guarantees the existence of a specification for which the matrix elements of the isomorphic potential  are  everywhere real.

%{\color{blue}{[REVIEWER II "what the physical meaning of the f * f 
%isomorphic matrix potential is": In the text above, we extend the logic of the adiabatic imaginary-time path-integral methods to the general multi-level cases. Recently, to understand the solid fundamental of those imaginary-time based methods, the connection between linearized real-time path-integral dynamics and the imaginary-time path-integral methods attracted the attention of the community (cite Althorpe, Hele, JCP, 2015; JCP, 2015;). Similarly, the connection between the exact non-adiabatic real-time propagator and our generalization would also be an interesting prospect to investigate. However at this stage, we have the confidence to draw the conclusion that our strategy captures the correct dynamics at certain limits and serves as a cornerstone for the future developments of the imaginary-time based dynamic methods.] [Maybe this point should be addressed after th summary?]}}

\section{Applications}
The isomorphic Hamiltonian can be used to incorporate nuclear quantum effects in any MQC simulation.
To illustrate this, present applications in which the isomorphic Hamiltonian is combined with either quantum-classical Liouville equation (QCLE) or fewest-switches surface hopping non-adiabatic dynamics.  Below, we briefly summarize the equations of motion associated with these two MQC methods.

\subsection{QCLE Dynamics}
\label{Sec:Methods:QCLE}

The time evolution of a general operator in a multi-level system according to QCLE dynamics is given by\cite{kapral1999mixed, nielsen2000mixed, nielsen2001statistical}
\begin{equation}
\label{qcle0}
\frac{\partial \hat{O}^{\rm W}(x, p, t)}{\partial t} = \hat{\mathcal{L}} \hat{O}^{\rm W}(x, p, t),
\end{equation}
where
\begin{equation}
\label{qcle1}
 \hat{\mathcal{L}} = \frac{i}{\hbar}\left[\hat{\mathcal{H}}, \bullet \right] - \frac12 \left( \left\{ \hat{\mathcal{H}},\bullet \right\}  - 
  \left\{ \bullet, \hat{\mathcal{H}}\right\}
  \right).
\end{equation}
In these equations, $\hat{O}^{\rm W}(x, p, t)$ is an $f\times f$ matrix  that corresponds to the partial Wigner distribution for a given operator with respect to a subset of the degrees of freedom,\cite{wigner1932quantum, hillery1997distribution} and $\hat{\mathcal{H}}$ is a generic Hamiltonian in the diabatic representation.

Our motivation for using the QCLE approach is to obtain a MQC limit in which the electronic dynamics evolves quantum mechanically and the nuclear dynamics evolves classically.  Taking the limit of small $\hbar$, the partial Wigner distribution reduces to the MQC phase-space distribution $\hat{O}$, such that 
the QCLE dynamics retains the same form, except that 
\begin{equation}
\label{qcle2}
\frac{\partial \hat{O}(x, p, t)}{\partial t} = \hat{\mathcal{L}} \hat{O}(x, p, t).
\end{equation}
Eqs.~\ref{qcle1} and \ref{qcle2}  thus cleanly define a MQC limit, where
the first term 
in the RHS of Eq.~\ref{qcle1}  describes the quantum evolution of the electronic states via the commutator, and the second term describes both the classical evolution of the nuclear coordinates and the back-reaction to the quantum subsystem via the symmetrized Poisson bracket.

Having taken the classical limit for the nuclei, the Kubo-transformed position-autocorrelation  function 
\begin{equation} 
\label{Kuboexact}
\tilde{c}_{xx} (t) 
=\frac{1}{\beta Q} \int_0^\beta d\lambda\ \text{tr} \left[ e^{-(\beta-\lambda) \hat{H}} \, \hat{x} \, e^{-\lambda \hat{H}}  \, \hat{x}(t) \right]
\end{equation}
becomes 
\begin{equation} 
\tilde{c}_{xx} (t) 
= \int \frac{dx \, dp}{2\pi\hbar }  \, {\rm tr_e} \left[ x e^{-\hat{\mathcal{L}}t}  \left(e^{-\beta\hat{\mathcal{H} }} \, x\right)  \right], \label{qclecorr2}
\end{equation}
where we have taken advantage of  time-reversal symmetry 
to  ensure that the time-evolved distribution in Eq.~\ref{qclecorr2} is conveniently numerically evaluated.

%\newpage
In this study, we consider the correlation function in Eq.~\ref{qclecorr2},  with the nuclei classically evolved either with respect to  the physical Hamiltonian ($\hat{\mathcal{H}}=\hat{H}$, where $\hat{H}$ is given in Eq.~\ref{systemnon}) or with respect to the CMD version of the isomorphic Hamiltonian ($\hat{\mathcal{H}}=\hat{H}^{\rm iso}_{\rm c}$, where $\hat{H}^{\rm iso}_{\rm c}$ is given in Eq.~\ref{centroidisoH}).
The resulting dynamics is used to study two- and three-level systems with a single nuclear degree of freedom.  Specifically, we investigate a two-level system comprised of shifted quartic oscillators with constant potential coupling, as well as a three-level system comprised of shifted harmonic oscillators with constant potential coupling.

The equations of motion in Eqs.~\ref{qcle1} and \ref{qcle2} are evolved exactly on a numerical grid, using the interaction picture with  Heisenberg evolution applied to the quantum subsystem; the resulting time-evolution  is both numerically stable and avoids additional approximations to the QCLE dynamics, such as the momentum-jump approximation.\cite{mac2002surface} 
The midpoint finite-difference method\cite{press2007numerical} is used to integrate the partial differential equations. 
We employ a numerical grid that spans the range of positions for which the classical Boltzmann probability density exceeds $10^{-12}$, 257 grid points in both $x$ and $p$ directions, and an integration timestep of $2.5\times 10^{-4}$ a.u.
The  matrix elements of the isomorphic potential, $\bar{V}_{i}^{\rm iso}$ and $\bar{K}_{ij}^{\rm iso}$, are sampled to convergence using path-integral Monte Carlo with $16\beta$ ring-polymer beads.

In the Results section, for comparison with the approximate QCLE dynamics described by Eqs.~\ref{qcle1} and \ref{qcle2}, 
we additionally obtain numerically exact quantum mechanical results by propagating the Schrodinger equation in the discrete variable representation (DVR)\cite{marston1989fourier, colbert1992novel} on a grid.
As is necessary, we confirm that the DVR results are identical to the QCLE dynamics in the high-temperature limit.  Additionally, for any temperature, we confirm that the DVR results are identical to the QCLE dynamics  for the case of a two-level system comprised of linearly coupled harmonic oscillators  when $\hat{\mathcal{H}}=\hat{H}$ and the dynamics is initialized from the multi-level partial Wigner phase-space distribution (Appendix \ref{appendixQCLE}).\cite{kapral1999mixed, nielsen2000mixed}

\subsection{Surface Hopping Dynamics}
\label{sec:methods:sh}

Consider a generic $f$-level system with $d$ nuclear degrees of freedom and diabatic Hamiltonian
\begin{equation}
\label{hopH}
\hat{\mathcal{H}}=\frac12 \sum_{j=1}^d m_j \dot{y}_j+\hat{\mathcal{V}}({\bf y}),
\end{equation}
where $\mathcal{V}({\bf y})$ is the diabatic potential energy matrix that depends on the nuclear positions, ${\bf y}=\{y_1,\ldots,y_d\}$, and $m_j$ is the mass of the $j^{\rm th}$ degree of freedom.
In  fewest-switches surface hopping,\cite{tully1990molecular} quantum evolution of the electronic wavefunction $\psi(\bf y, t)$ along a given trajectory obeys
\begin{equation}
\label{FSSH-elec}
i\hbar \frac{\partial}{\partial t}\psi({\bf y}, t) = \hat{\mathcal{V}}({\bf y}) \psi({\bf y}, t),
\end{equation}
and classical evolution of the nuclear coordinates obeys
\begin{equation}
\label{FSSH-nuc}
m_j \ddot{y}_j=-\frac{\partial}{\partial y_j}\ \mathcal{E}_k({\bf y}), 
\end{equation}
where $\mathcal{E}_k$ is the $k^{\rm th}$ adiabatic Born-Oppenheimer surface obtained by diagonalizing the diabatic potential matrix. 
The nuclear trajectory evolves along a particular Born-Oppenheimer surface, subject to stochastic hops to other surfaces with probability
%\begin{equation}
%\label{hopprob}
%{\color{red}\sout{p_{kl}=\textrm{max}\left\{\frac{1}{a_{kk}}\frac{da_{ll}}{dt}\Delta t, 0\right\}}}
%\end{equation} 
\begin{equation}
\label{hopprob}
p_{kl}=\textrm{max}\left\{-\frac{2}{a_{kk}} \text{Re}((d_{lk} \cdot v) a_{kl}) \Delta t, 0\right\}
\end{equation} 
where $a_{kl}$ is the element of the electronic density matrix in the adiabatic representation, $(d_{lk} \cdot v)$ is the inner product of the first-derivative non-adiabatic coupling with the nuclear velocity vector, and $\Delta t$ is the integration timestep. 
%{\color{blue}{[REVIEW I, POINT II: I suggest to rewrite the formula to the general multi-state case@Tully, JCP, 1990.]}}
During hopping events, the total energy associated with the Hamiltonian in Eq.~\ref{hopH} is conserved by modifying the component of the velocity along the  non-adiabatic coupling vector that connects the two surfaces; hops are forbidden if there is insufficient velocity in this component to ensure energy conservation.  We implement forbidden hops without momentum reversal,\cite{jain2015surface, hammes1994proton} and we neglect decoherence corrections,\cite{subotnik2011new, prezhdo1998relationship} although either could easily be implemented in the current context.

In this study, we consider various implementations of fewest-switches surface hopping in a two-level gas-phase scattering system that is a function of a single nuclear coordinate:

{\it (i)}  For the standard case of surface-hopping with classical nuclei (hereafter referred to as SH-classical), we employ Eqs.~\ref{hopH}-\ref{hopprob} using the physical Hamiltonian ($\hat{\mathcal{H}}=\hat{H}$,  given in Eq.~\ref{systemnon}) which includes the physical diabatic potential matrix ($\hat{\mathcal{V}}=\hat{V}$) as a function of the single nuclear coordinate, such that ${\bf y}=x$. 
  
{\it (ii)} To quantize the nuclei in the surface hopping dynamics with the CMD version of the isomorphic Hamiltonian (referred to as SH-C-iso), we employ Eqs.~\ref{hopH}-\ref{hopprob} using $\hat{\mathcal{H}}=\hat{H}^{\rm iso}_{\rm c}$ (given in Eq.~\ref{centroidisoH}), which includes the CMD version of the  diabatic potential matrix ($\hat{\mathcal{V}}=\hat{V}^{\text{iso}}_{\rm c}$, given in Eq.~\ref{Visoc})  as a function of the centroid nuclear coordinate, such that ${\bf y}=\bar{x}$.

{\it (iii)} To quantize the nuclei in the surface hopping dynamics with the RPMD version of the isomorphic Hamiltonian (referred to as SH-RP-iso), we employ Eqs.~\ref{hopH}-\ref{hopprob} using $\hat{\mathcal{H}}=\hat{H}^{\rm iso}_{n}$ (given in Eq.~\ref{isoHam}), which includes the RPMD version of the  diabatic potential matrix ($\hat{\mathcal{V}}=U_{\rm spr} + \hat{V}^{\text{iso}}$, given in Eqs.~\ref{Uspr} and \ref{Viso}, respectively)  as a function of the ring-polymer coordinates, such that ${\bf y}={\bf x}$.  

{\it (iv)} Finally, for comparison with an earlier effort to combine RPMD with surface hopping, we also employ the method described in Ref.~\onlinecite{shushkov2012ring} using the ``bead-approximation" defined therein; this method is referred to as SH-RP-nokinks, since it neglects the  contribution of the ``kinked" ring-polymer configurations that span multiple diabatic surfaces, such that Eq.~\ref{isox} is not obeyed and the quantum Boltzmann statistics are approximated.

Note that for all surface-hopping calculations reported here, the dynamics is run in a representation for which the number of electronic states is the same as for the physical system.  For  results obtained using the various versions of the isomorphic Hamiltonian, the surface-hopping dynamics involves transitions between the adiabatic  potential surfaces obtained by diagonalizing the isomorphic diabatic potential energy matrix.  

Following the implementation in Ref.~~\onlinecite{shushkov2012ring}, Eq.~\ref{FSSH-elec} is evolved in the interaction representation using a fourth-order Runge-Kutta integrator,\cite{press2007numerical}
and Eq.~\ref{FSSH-nuc} is evolved using the velocity Verlet algorithm.\cite{frenkel2001understanding}  As in previous
RPMD simulations, each timestep for the nuclear degrees of freedom
involves separate coordinate updates due to forces arising
from the adiabatic potential and due to exact evolution of the
purely harmonic portion.\cite{miller2005quantum, habershon2013ring}
Matrix elements of the centroid isomorphic potential, $\bar{V}_{i}^{\rm iso}$ and $\bar{K}_{ij}^{\rm iso}$, are sampled to convergence using path-integral Monte Carlo with either $8\beta$ ring-polymer beads (for $\beta\le 9$) or $24\beta$ ring-polymer beads (for $\beta > 9$); the larger number of ring-polymer beads was found to be more important for improving statistical sampling of the centroid potential surfaces than for converging the path-integral discretization.
The SH-RP-iso results were likewise performed using $8\beta$ ring-polymer beads. %{\color{blue}{Inset: SH-C-iso, to keep the potential smooth, 24 $\beta$ beads is used.}}
For all cases, Eq.~\ref{FSSH-nuc} is integrated with a timestep of $10^{-4}$ a.u.
Thermal rates in this study are calculated via Boltzmann averaging of the microcanonical reactive probabilities, initializing trajectories outside of the interaction region with a momentum range for which the ratio of the corresponding Boltzmann-weighted microcanonical reactive probability to the total thermal rate is greater than $10^{-8}$ a.u.
For the SH-classical and  SH-C-iso calculations, for which the microcanonical reactive probability changes abruptly at the threshold energy, we discretize this momentum interval at a resolution of $0.01$ a.u; for the SH-RP-iso and  SH-RP-nokinks calculations, we use a discretization of $0.05$ a.u.  
The microcanonical reactive probabilities are calculated using from 10$^4$ to 10$^5$ trajectories.

% this range is discretized with a momentum interval of $0.05$ a.u., {\color{blue}{for SH-C-iso, a momentum interval of $0.01$ a.u is applied to capture the stepsize shape of the scattering probability.}} and each microcanonical reactive probability is calculated using 10$^4$ trajectories. {\color{blue}{Inset: for SH-RP-nokinks and SH-RP-iso, more trajectories for low momenta are generated to reduce the uncertainty from sampling low scattering probability.}} 

In the results section, for comparison with the various surface-hopping implementations, 
we additionally obtain numerically exact quantum mechanical results via wavepacket propagation, using the split-operator Fourier transform method of Feit and Fleck\cite{feit1982solution} extended to multiple potential energy surfaces. A wavepacket was initialized in the asymptotic reactant region and evolved forward in time until the scattering event was completed. An absorbing potential was placed in the asymptotic reactant region that eliminated the reflected portion of the scattered wavepacket, while the transmitted component was projected out in the asymptotic product region. The scattering amplitudes were calculated by Fourier transform of the transmitted fraction of the wavepacket, properly normalized, and the squared modulus of the scattering amplitudes is numerically integrated to obtain the quantum rates.

To illustrate the full details of our implementation of the SH-RP-iso method, we have provided an example program online. \cite{code}

\section{RESULTS}

We now present numerical results for two possible combinations of the new path-integral isomorphic Hamiltonian with MQC methods.  First, to investigate a well-defined limit for MQC non-adiabatic dynamics in combination with the isomorphic Hamiltonian, we employ the QCLE method, considering both a two-level system of coupled quartic oscillators and a three-level system  involving a donor-bridge-acceptor model.  Then, to investigate a broadly applicable combination of MQC non-adiabatic dynamics with the isomorphic Hamiltonian, we employ fewest-switches surface hopping to study a model for state-resolved gas-phase reactive scattering. 
Unless otherwise specified, quantities are reported in atomic units, and we employ a nuclear mass of $m=1$.

\subsection{QCLE Dynamics}
\subsubsection{Two-level system: Coupled quartic oscillators}

\begin{figure*}
\includegraphics[width=16.8cm]{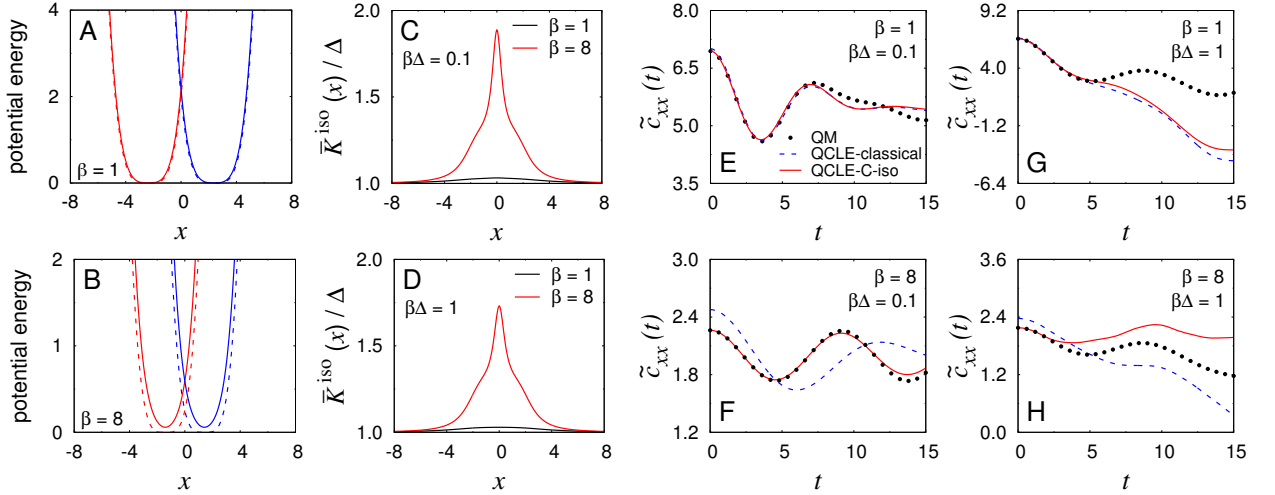}
\caption{\label{fig_1} 
{\bf(A,B)} Diagonal potential energy matrix elements for the coupled quartic oscillator system at high temperature ($\beta=1$, panel A) and  low temperature ($\beta=8$, panel B). Matrix elements for the physical potential $V_i(x)$ and for the CMD version of the isomorphic potential $\bar{V}_{i}^{\rm iso}(\bar{x})$ are shown in dashed and solid lines, respectively.  Matrix elements for diabats 1 and 2 are shown in blue and red, respectively.
{\bf(C,D)} The  off-diagonal matrix element of the CMD version of the isomorphic potential, $\bar{K}_{12}^{\rm iso}$, normalized by the off-diagonal coupling $\Delta$ in the physical potential, for weak coupling ($\beta\Delta=0.1$, panel C) and for intermediate coupling ($\beta\Delta=1$, panel D). High-temperature ($\beta=1$) and low-temperature ($\beta=8$) results are shown in black and red, respectively.
{\bf(E-H)} Kubo-transformed position-autocorrelation functions obtained using exact quantum mechanics (QM; black, dots), QCLE dynamics with classical nuclei (QCLE-classical; blue, dashed), and QCLE dynamics with nuclei quantized via the CMD version of the isomorphic Hamiltonian (QCLE-C-iso; red, solid).
Results are presented for weak coupling and high temperature ($\beta\Delta=0.1$, $\beta=1$; panel E),  weak coupling and low temperature ($\beta\Delta=0.1$, $\beta=8$; panel F),  intermediate coupling and high temperature ($\beta\Delta=1$, $\beta=1$; panel G), and  intermediate coupling and low temperature ($\beta\Delta=1$, $\beta=8$; panel H).
 }
\end{figure*}

We begin by considering a two-level system involving a single nuclear coordinate, for which the physical potential energy matrix, $\hat{V}(x)$, % is
is comprised of diagonal elements that are strongly anharmonic quartic oscillators, $V_{1}(x) =  (x+x_0)^4/16$ and $V_{2}(x) =  (x-x_0)^4/16$, and the off-diagonal elements, $K_{12}(x) = \Delta$, are constant.  The lateral shift of the potentials is $x_0=(32/\beta)^{1/4}$, such that the activation energy associated with the crossing of the diabats is consistently $2/\beta$.
%{\color{blue}{The model systems are picked in this way such that we can investigate the transition dynamics between wells rather than that within a well, even in the lower temperature case. In other words, the parameters  in the physical potential are dependent on the temperature, but they are just constants to make sure the barrier height do not change relative to the thermal temperature.}}
 In studying this system, we will consider \textit{(i)} numerically exact quantum dynamics,  \textit{(ii)} the classical nuclear limit in which the QCLE dynamics is  run using the physical Hamiltonian, $\hat{H}(x)$, and \textit{(iii)} the case of quantized nuclei in which the QCLE dynamics is run using the CMD version of the isomorphic Hamiltonian, $\hat{H}^{\rm iso}_{\rm c}$.  Methodological and computational details are provided in Section~\ref{Sec:Methods:QCLE}.

Figs.~\ref{fig_1}A-D illustrate  
the matrix elements of the CMD version  of the isomorphic potential, $\hat{V}^{\rm iso}_{\rm c}$ (Eq.~\ref{Visoc}).  In solid lines, panels A and B present the diagonal elements of the isomorphic potential, $\bar{V}_{1}^{\rm iso}(\bar{x})$ and $\bar{V}_{2}^{\rm iso}(\bar{x})$, at high and low temperature, with the physical diabatic potentials $V_{1}(x)$ and $V_{2}(x)$ shown in dashed lines for comparison.  Given that these isomorphic potential matrix elements are identical to the  CMD potentials of mean force for the two diabats, they exhibit the familiar features of converging to the physical potential at  high temperature  (Fig.~\ref{fig_1}A) and exhibiting larger nuclear quantization effects at low temperature (Fig.~\ref{fig_1}B). 

For weak coupling ($\beta\Delta=0.1$) and intermediate coupling ($\beta\Delta=1$),  Figs.~\ref{fig_1}C and D respectively present the off-diagonal matrix elements of the isomorphic potential, $\bar{K}_{12}^{\rm iso}$,   at both low (red) and high (black) temperature.  Unlike the coupling in the physical potential for this model, $\Delta$, the coupling in the isomorphic potential is position dependent, reflecting the changing thermal probability of kinked ring-polymer configurations at different nuclear configurations.  In all cases, the inclusion of nuclear quantization via exact path-integral statistics leads to an increase in the effective coupling between the two diabatic surfaces in the vicinity of the diabatic crossing ($x=0$), with more pronounced effects at lower temperature.

Figs.~\ref{fig_1}E and F present results for the Kubo-transformed position-autocorrelation function (Eq.~\ref{Kuboexact}) in the weak-coupling regime ($\beta\Delta=0.1$) at high and low temperature, respectively. At the higher temperature (Fig.~\ref{fig_1}E), there is little difference in the QCLE dynamics obtained with classical nuclei (QCLE-classical; blue, dashed) versus with nuclei quantized via the CMD version of the isomorphic Hamiltonian (QCLE-C-iso; red, solid), and both implementations of QCLE are in good agreement with exact quantum mechanics (black, dots) due to the small role of nuclear quantum effects.  At low temperatures, however, substantial nuclear quantum effects emerge, as evidenced by the difference between the blue and black curves in Fig.~\ref{fig_1}F.  In this low-temperature case, the QCLE-C-iso dynamics exhibit substantial improvement, recovering the exact quantum result at $t=0$ as a necessary consequence of the  path-integral statistics and showing better agreement with the quantum mechanical period of oscillation.

Finally, Figs.~\ref{fig_1}G and H present results for the Kubo-transformed position-autocorrelation function in the intermediate-coupling regime ($\beta\Delta=1$) at high and low temperature, respectively.  As before, at high temperature (Fig.~\ref{fig_1}G),  the QCLE-classical dynamics differs little from the QCLE-C-iso dynamics; however, both differ substantially from the exact quantum result at longer times.  At low temperature (Fig.~\ref{fig_1}H), even larger differences are observed.  As is necessary,  QCLE-C-iso recovers the exact quantum result at short times, but it deviates from both QCLE-classical and exact quantum results at longer times.

As is familiar from standard CMD and RPMD in one-level systems,\cite{craig2004quantum, cao1994formulation2} 
the results in  Figs.~\ref{fig_1}E-H highlight that the  newly introduced isomorphic Hamiltonian provides a means of exactly incorporating the statistical effects of nuclear quantization while only approximately including the dynamical effects.  Moreover, the dynamics obtained from the isomorphic Hamiltonian will reflect the particular shortcomings of the employed MQC method - in this case,  QCLE initialized with the MQC phase-space distribution.  In Appendix D, we illustrate that a leading source of error for the QCLE-C-iso results in Figs.~\ref{fig_1}E-H is non-preservation of the MQC phase-space distribution in the QCLE dynamics at lower temperatures, where the MQC phase-space  distribution differs substantially from the partial Wigner distribution.

\subsubsection{Three-level system: Donor-Bridge-Acceptor model}

\begin{figure}
\includegraphics[width=8.4cm]{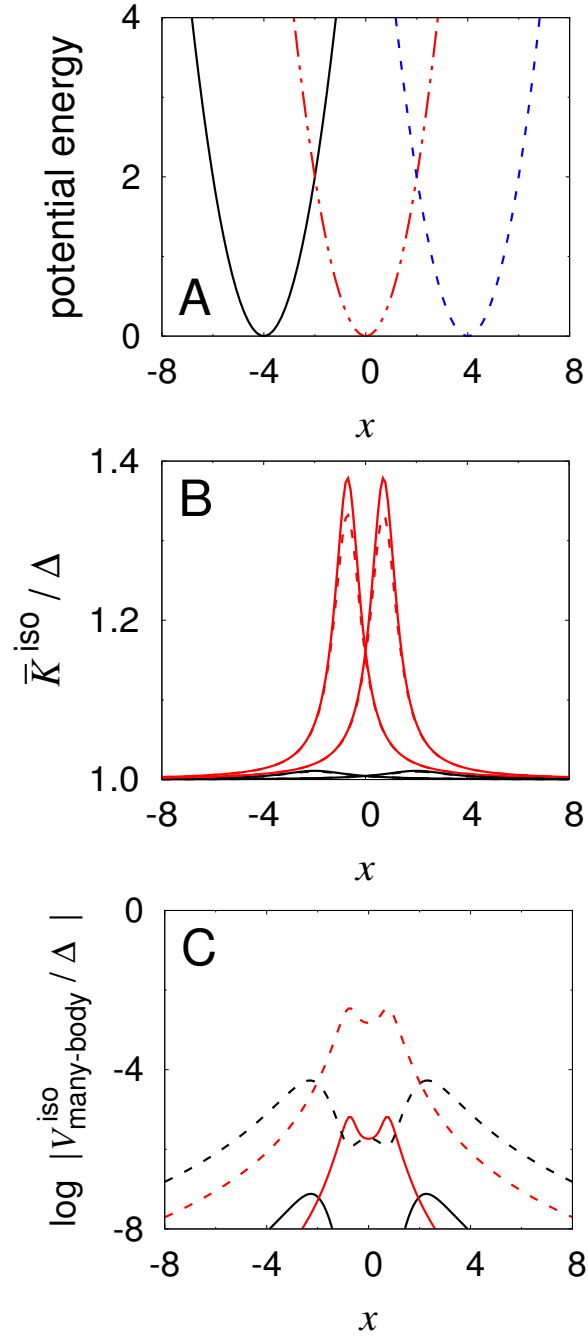}
\caption{\label{fig_2} 
{\bf(A)} Diagonal matrix elements of the physical potential for the three-level donor-bridge-acceptor system  with $\beta=1$. 
{\bf(B)} Off-diagonal matrix elements of the CMD version of the isomorphic potential, $\bar{K}^{\rm iso}$, normalized by $\Delta$. 
{\bf(C)} Many-body contribution to the isomorphic potential of the three-level system, $\bar{V}^{\text{iso}}_{\textrm{many-body}}$, normalized by $\Delta$. 
Results are presented for weak coupling and high temperature ($\beta\Delta=0.1$, $\beta=1$; black, solid),  weak coupling and low temperature ($\beta\Delta=0.1$, $\beta=8$; red, solid),  intermediate coupling and high temperature ($\beta\Delta=1$, $\beta=1$; black, dashed), and  intermediate coupling and low temperature ($\beta\Delta=1$, $\beta=8$; red, dashed). In panel B, the high-temperature results (black lines) are graphically indistinguishable.
}
\end{figure}

For systems with more than two levels, a many-body correction appears in the  isomorphic potential to ensure exact Boltzmann statistics ($V^{\text{iso}}_{\textrm{many-body}}$ in Eq.~\ref{Viso} and $\bar{V}^{\text{iso}}_{\textrm{many-body}}$ in Eq.~\ref{Visoc}). 
To investigate the nature of this many-body term, 
we  consider a previously studied model for a three-level donor-bridge-acceptor system.\cite{jang2001nonadiabatic}
For this system, the physical potential energy, $\hat{V}(x)$,
is comprised of diagonal elements that are harmonic oscillators ($V_{1}(x) =  (x+x_0)^2/2$,  $V_{2}(x) =  x^2/2$, and $V_{3}(x) =  (x-x_0)^2/2$), and the off-diagonal elements are constant ($K_{12}(x) = K_{23}(x) = \Delta$, $K_{13}(x) = 0$).  The lateral shift of the potentials is $x_0=4/\beta^{1/2}$, such that the activation energy associated with the crossing of the diabats is $2/\beta$.  For the case of $\beta=1$, the diagonal elements of the physical potential are shown in Fig.~\ref{fig_2}A.

Upon computing the matrix elements for the CMD version of the isomorphic potential, $\hat{V}^{\rm iso}_{\rm c}$  (Eq.~\ref{Visoc}), it is found that the diagonal (not shown) and off-diagonal (Fig.~\ref{fig_2}B) contributions to the two-body isomorphic potential (Eq.~\ref{Visoc2b}) are qualitatively similar to those illustrated in  Figs.~\ref{fig_1}A-D.  The many-body contribution to the isomorphic potential of the three-level system, $\bar{V}^{\text{iso}}_{\textrm{many-body}}$, is plotted in Fig.~\ref{fig_2}C, divided by $\Delta$ to illustrate the magnitude of this many-body term in comparison to the two-body potential coupling.  As is clear from the log-scale in Fig.~\ref{fig_2}C, we find in all studied cases that the  many-body contribution is negligible in comparison to the two-body coupling between the electronic states.  As a result, the dynamics for this system exhibits very little three-body character, and the computed time correlation functions (not shown) exhibit the qualitative features of those discussed in Figs.~\ref{fig_1}E-H.
We thus find that the isomorphic Hamiltonian can be straightforwardly applied in multi-level systems and that, at least for the three-level system studied here, the many-body contribution to the isomorphic potential plays a minor role.

\subsection{Surface-hopping dynamics}

\begin{figure}
\includegraphics[width=8.4cm]{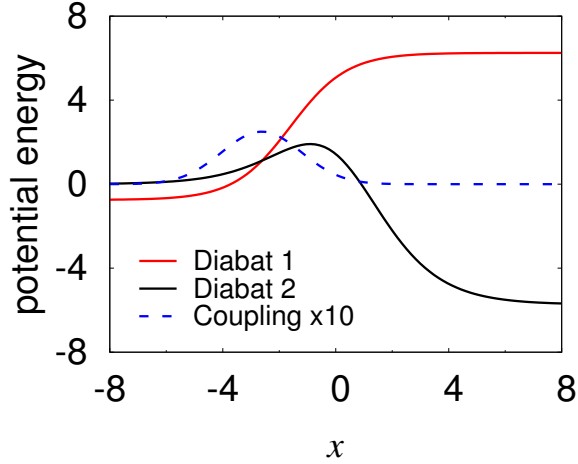}
\caption{\label{fig_3} 
Matrix elements of the physical potential for the two-level reactive scattering system, including diabat 1 (red), diabat 2 (black), and the off-diagonal coupling (blue, dashed,  with 10-fold magnification).
}
\end{figure}

\begin{table}
\caption{\label{tab:table1} Parameter values for the  physical potential of the two-level reactive scattering system, given in Eq.~\ref{scatpot}.}
\begin{ruledtabular}
\begin{tabular}{cccc}
Parameter&\mbox{Value}&Parameter&\mbox{Value}\\
\hline
$A_1$ & $7$ & $a_1$ & $1$ \\
$A_2$ & $-18/\pi$ & $a_2$ & $\sqrt{3\pi}/4$ \\
$A_3$ & $0.25$ & $a_3$ & $0.25$ \\
$B_1$ & $-0.75$ & $x_1$ & $-1.6$ \\
$B_2$ & $54 / \pi$ & $x_3$ & $-2.625$ \\
\end{tabular}
\end{ruledtabular}
\end{table}

\begin{figure}
\includegraphics[width=8.4cm]{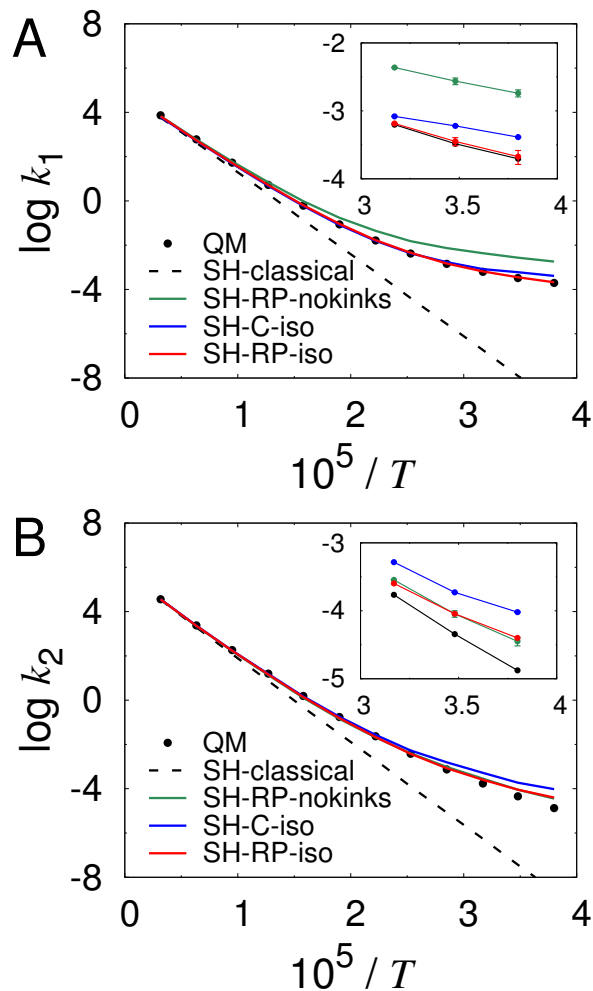}
\caption{\label{fig_3new} 
State-to-state thermal reaction rates as a function of temperature, obtained using surface hopping with classical nuclei (SH-classical; black, dashed) and with  nuclei quantized via the SH-RP-nokinks (green), SH-C-iso (blue), and SH-RP-iso (red) methods, as well as with exact quantum mechanics (black, dots). 
{\bf(A)} The rate ($k_1$) for the  channel that enters on diabat 1 and exits on diabat 2.
{\bf(B)}
The rate ($k_2$) for the  channel that enters on diabat 2 and exits on diabat 2.
The insets expand the axes in the low-temperature region. Unless explicitly shown, the error bars are smaller than the size of the plotted circles. For the inset of panel (B), the SH-RP-iso and SH-RP-nokinks results are within the statistical error at all temperatures. 
Both temperature and the reaction rate are reported in SI units.  
}
\end{figure}

We finally consider the state-to-state reactive scattering in a two-level model for a gas-phase system with a single nuclear degree of freedom.  The physical potential for this system is given by matrix elements
\begin{eqnarray}
\label{scatpot}
V_{1}(x)&=&\frac{A_1}{1+e^{-a_1\left(x-x_1\right)}}+B_1\nonumber\\
V_{2}(x)&=&\frac{A_2}{1+e^{-a_2x}}+\frac{B_2}{4\ {\rm cosh}^2\left(\frac{a_2x}{2}\right)}\\
K_{12}(x)&=&A_3e^{-a_3\left(x-x_3\right)^2}\nonumber
\end{eqnarray}
with parameters given in Table~\ref{tab:table1}. Both the diagonal and off-diagonal potential matrix elements are plotted in Fig.~\ref{fig_3}, with reactants at $x\rightarrow-\infty$ and products at $x\rightarrow\infty$.  
The basic features of this model resemble the F+H$_2$ co-linear reaction, exhibiting both endothermal and exothermal reactive channels.  We consider the thermal reaction rate $k_1$ for the channel that  enters on diabatic state 1 and exits on diabatic state 2, as well as the 
 thermal reaction rate $k_2$ for the channel that  enters on diabatic state 2 and exits on diabatic state 2.
 The state-to-state thermal reaction rates are calculated using methods that include \textit{(i)} numerically exact quantum dynamics,  \textit{(ii)}  surface hopping with classical nuclei (SH-classical), \textit{(iii)}  surface hopping with nuclei quantized via the ring-polymer surface hopping method in Ref.~\onlinecite{shushkov2012ring} that approximates the path-integral statistical distribution (SH-RP-nokinks), 
  \textit{(iv)}  surface hopping with nuclei quantized via the CMD version of the isomorphic Hamiltonian (SH-C-iso),
 and 
   \textit{(v)}  surface hopping with nuclei quantized via the RPMD version of the isomorphic Hamiltonian (SH-RP-iso).
Both the SH-C-iso and SH-RP-iso methods are newly presented in this work.  Results were also obtained using classical Ehrenfest dynamics \cite{ehrenfest1933phase}  but are excluded due to their  poor quality for this model.
 Computational details are provided in Section \ref{sec:methods:sh}, and an example program that runs the SH-RP-iso trajectories for the system studied here is provided online. \cite{code}  

 Figure \ref{fig_3new}A presents results for the thermal reaction rate $k_1$ obtained using the various methods as a function of reciprocal temperature, with the inset providing an expanded view of the lowest-temperature results.  The large differences between the exact quantum and SH-classical results at low temperature illustrate the strong role of nuclear quantum effects. %{\color{blue}{Though it is not presented here, Ehrenfest is also terrible for the scattering process (maybe cite Tully, JCP, 2006?) while MF-RPMD would be unable to describe the state-to-state processes since it would thermally average over all states. Next,}}
 Although the SH-RP-nokinks method qualitatively recovers the effect of nuclear tunneling in this process, it  overestimates the thermal reaction rate at low temperatures by at least an order of magnitude (see inset).  Since SH-RP-nokinks neglects  ring-polymer configurations that span the two electronic surfaces, it underestimates the role of the low-lying excited state in suppressing nuclear tunneling; similar errors are observed when standard RPMD on the lower adiabatic surface is used to approximate tunneling through an avoided crossing (see Fig.~2 of Ref.~\onlinecite{kretchmer2013direct}).
 It is clear that both the SH-C-iso and SH-RP-iso results in  Fig.~\ref{fig_3new}A are in better agreement with the exact quantum results, with the RPMD version of the isomorphic Hamiltonian leading to particularly accurate results.
 
  Figure \ref{fig_3new}B presents the corresponding results for the thermal reaction rate $k_2$.  Again, large nuclear quantum effects at low temperature are indicated by the difference between the exact quantum and SH-classical results.
  The inset reveals that for this reactive channel, the SH-C-iso method exhibits the largest errors among the quantized surface hopping methods, 
overestimating the reaction rate by an order of magnitude in the deep-tunneling regime ($\beta>\beta_{\rm c}\approx 8$ for diabat 2). 
This result illustrates a well-known shortcoming of CMD for deep-tunneling across asymmetric barriers,\cite{jang1999modification}
which is the precise nature of the reaction channel associated with $k_2$.
For this process, the SH-RP-iso and SH-RP-nokinks are graphically indistinguishable and are in good agreement with the exact quantum results.

We note that this simple model for a gas-phase scattering reaction reveals a significant shortcoming of both the SH-RP-nokinks and  CMD-based methods for describing non-adiabatic chemical dynamics.
Surface hopping combined with the RPMD version of the isomorphic Hamiltonian (SH-RP-iso) avoids these pitfalls and provides the best accuracy for both reactive channels at all temperatures.

%\newpage
\section{SUMMARY}

The current work strives to decouple the methodological challenge of describing electronically non-adiabatic dynamics  from that of 
describing  nuclear quantization.  
For a general physical system with multiple electronic energy levels, we derive a corresponding isomorphic Hamiltonian, such that Boltzmann sampling of the isomorphic Hamiltonian with classical nuclear degrees of freedom yields the exact quantum Boltzmann distribution for the original physical system.    The key  advantage of this isomorphic Hamiltonian is that it can  be combined with existing mixed quantum-classical (MQC)   methods for non-adiabatic dynamics, allowing for the straightforward inclusion of nuclear quantum effects.

The isomorphic Hamiltonian is presented in two versions, one of which recovers  standard ring-polymer molecular dynamics (RPMD) in the limit of a single electronic surface, and the other that recovers standard centroid molecular dynamics (CMD).  Numerical results are presented using both the RPMD and CMD versions of the isomorphic Hamiltonian, in combination with either  fewest-switches surface hopping  or the  quantum-classical Liouville equation (QCLE) descriptions of MQC non-adiabatic dynamics.   Investigation of a simple model for non-adiabatic gas-phase scattering reveals that a particularly promising  approach is to combine surface-hopping dynamics with the RPMD version of the isomorphic Hamiltonian (i.e., the SH-RP-iso method), which exhibits the best accuracy among the studied methods for two different reactive channels at all temperatures.

Future work will include applications of the isomorphic Hamiltonian 
to explore the role of nuclear quantum effects in the non-adiabatic dynamics of complex systems.  
Methodological extensions of the current work are also of interest, including alternative specification of the matrix elements of the isomorphic Hamiltonian (as discussed in Section \ref{sec:multilevel}), and combination of the isomorphic Hamiltonian with  other MQC  methods for describing non-adiabatic dynamics.  Also of interest  are dimensionality-reduction strategies
based on generalization of the isomorphic potential energy  in Eq.~\ref{Viso}  to describe the correlated dynamics of a local subset of electronic states embedded in a  mean-field treatment of the environment (akin to quantum embedding strategies for electronic structure\cite{manby2012simple}).

\begin{acknowledgments}
We acknowledge support from the
Office of Naval Research  under Award Number N00014-10-1-0884
and the 
Air Force Office of Scientific Research under Award Number FA9550-17-1-0102.
Additionally, 
P.S. acknowledges a German Research Foundation (DFG) Postdoctoral Fellowship, and 
T.F.M. acknowledges
a Camille Dreyfus Teacher-Scholar Award.
Computational
resources were provided by the National Energy Research Scientific
Computing Center, which is supported by the Office of Science of the US
Department of Energy under Contract No. DE-AC02-05CH11231.
\end{acknowledgments}

\appendix

\section{Equivalent forms of the ring-polymer Hamiltonian}
The ring-polymer Hamiltonian is usually introduced \cite{craig2004quantum, habershon2013ring} 
 by writing the partition function as 
\begin{eqnarray}
\label{app:pf1}
Q=\lim_{n\rightarrow\infty} \left( 2\pi\hbar  \right)^{-n} \!\! \int \!d\textbf{x} \int \! d\textbf{p}\ e^{-\beta_n H_n(\textbf{x}, \textbf{p})},
\end{eqnarray}
where
\begin{equation} \label{app:rpmdH}
H_n = \sum_{\alpha=1}^n \frac{p_{\alpha}^2}{2m}  + n U_{\text{spr}}(\textbf{x}) + \sum_{\alpha=1}^n V(x_{\alpha}) 
\end{equation}
and $U_{\text{spr}}(\textbf{x})$ is defined in Eq.~\ref{Uspr}.
The RPMD equations of motion associated with this form of the  Hamiltonian are
\begin{eqnarray}
\label{app:rpmd_eom}
\dot{x}_{\alpha}&=&p_{\alpha}/m\\
\dot{p}_{\alpha}&=&m\omega_n^2\left(x_{(\alpha+1)}+x_{(\alpha-1)}-2x_{\alpha}\right)-\frac{\partial}{\partial x_{\alpha}} V\!\left(x_{\alpha}\right) \nonumber
\end{eqnarray}
or
\begin{equation}
\label{app:rpmd_eom1}
\ddot{x}_{\alpha}=\omega_n^2\left(x_{(\alpha+1)}+x_{(\alpha-1)}-2x_{\alpha}\right)-\frac1m \frac{\partial}{\partial x_{\alpha}} V\!\left(x_{\alpha}\right)
\end{equation}
for $\alpha=1,\ldots,n$, and 
the Lagrangian associated with this Hamiltonian is
\begin{equation}\label{app:rpmd_L}
\mathcal{L}=\sum_{\alpha=1}^n \frac12 m \dot{x}_{\alpha}^2  - n U_{\text{spr}}(\textbf{x}) - \sum_{\alpha=1}^n V(x_{\alpha}).
\end{equation}

Now, we introduce a new Lagrangian that is obtained by constant scaling of the original,
\begin{equation}\label{app:rpmd_L}
\mathcal{L}^{\rm iso} \equiv  \mathcal{L}/n,
\end{equation}
which yields the corresponding Hamiltonian
\begin{equation} \label{app:isoH}
H_n^{\rm iso} = \sum_{\alpha=1}^n \frac{(p^{\rm iso}_{\alpha})^2}{2m_n}  + U_{\text{spr}}(\textbf{x}) + \frac1n \sum_{\alpha=1}^n V(x_{\alpha}).
\end{equation}
The classical equations of motion associated with this Hamiltonian are
\begin{eqnarray}
\label{app:rpmd_eom}
\dot{x}_{\alpha}&=&p^{\rm iso}_{\alpha}/m_n\\
\dot{p}^{\rm iso}_{\alpha}&=&m_n\omega_n^2\left(x_{(\alpha+1)}+x_{(\alpha-1)}-2x_{\alpha}\right) - \frac1n \frac{\partial}{\partial x_{\alpha}} V\!\left(x_{\alpha}\right) \nonumber
\end{eqnarray}
or
\begin{equation}
\label{app:rpmd_eom2}
\ddot{x}_{\alpha}=\omega_n^2\left(x_{(\alpha+1)}+x_{(\alpha-1)}-2x_{\alpha}\right)-\frac1m \frac{\partial}{\partial x_{\alpha}} V\!\left(x_{\alpha}\right)
\end{equation}
Comparison of Eqs.~\ref{app:rpmd_eom1} and \ref{app:rpmd_eom2} confirms that since the two forms of the Hamiltonian (in Eqs.~\ref{app:rpmdH} and \ref{app:isoH}) are obtained from constant scaling of the same Lagrangian, they yield the same equations of motion.

Finally, we can rewrite the exponand in Eq.~\ref{app:pf1} as
\begin{eqnarray}
-\beta_n H_n 
&=&
-\beta \left[\frac1n \sum_{\alpha=1}^n \frac{p_{\alpha}^2}{2m}  +   U_{\text{spr}}(\textbf{x}) +  \frac1n \sum_{\alpha=1}^n V(x_{\alpha})\right] \nonumber\\
&=&
-\beta \left[ \sum_{\alpha=1}^n \frac12 m_n \dot{x}_{\alpha}^2  +   U_{\text{spr}}(\textbf{x}) +  \frac1n \sum_{\alpha=1}^n V(x_{\alpha})\right] \nonumber\\
&=&
-\beta \left[ \sum_{\alpha=1}^n \frac{(p^{\rm iso}_{\alpha})^2}{2m_n}  +   U_{\text{spr}}(\textbf{x}) +  \frac1n \sum_{\alpha=1}^n V(x_{\alpha})\right] \nonumber\\
&=&
-\beta H_n^{\rm iso}  \nonumber
\end{eqnarray}

We have thus shown that the partition function in Eq.~\ref{app:pf1} can equivalently be rewritten as
\begin{equation}
\label{app:pf2}
Q=\lim_{n\rightarrow\infty} \left(\frac{n}{2\pi \hbar}  \right)^{n} \!\! \int \!d\textbf{x} \int \! d\textbf{p}^{\rm iso} \ e^{-\beta H_n^{\rm iso}(\textbf{x}, \textbf{p}^{\rm iso})}
\end{equation}
and that the Hamiltonian in Eq.~\ref{app:isoH} yields the usual RPMD equations of motion.  In the main text, we  employ Eqs.~\ref{app:pf2} and \ref{app:isoH} for the partition function and the ring-polymer Hamiltonian, respectively, and for succinctness, we drop the superscript ``iso" in denoting the bead momenta.

\section{CMD version of the isomorphic Hamiltonian}

The 
 CMD version of the isomorphic Hamiltonian is
 \begin{equation}  
 \label{centroidisoH}
\hat{H}^{\rm iso}_{\rm c}(\bar{x}, \bar{p}) =  \frac{\bar{p}^2}{2m}  + \hat{V}^{\rm iso}_{\rm c}(\bar{x}),
\end{equation}
where
$\hat{V}^{\rm iso}_{\rm c}$ is the isomorphic potential energy  given by the $f\times f$ matrix that obeys
\begin{equation} 
\textrm{tr}_{\textrm{e}}\left[  e^{-\beta \hat{V}^{\rm iso}_{\rm c}(\bar{x}) } \right] \equiv \bar{\mu} (\bar{x}),
\end{equation}
\begin{equation} 
 \bar{\mu} (\bar{x})
 =   \lim_{n\rightarrow\infty} C  \int \!\!\! d\textbf{x} \ \delta (\bar{x}-\frac1n \!\sum_{\alpha} x_{\alpha})\, e^{-\beta U_{\text{spr}}(\textbf{x})} \, \mu(\textbf{x}),
\end{equation}
$C=\sqrt{n}\left( \frac{mn}{2\pi \beta \hbar^2} \right)^{(n-1)/2} $,
and $\mu(\textbf{x})$ is given by Eq.~\ref{mu}.
Following the logic of the main text, we obtain the centroid isomorphic potential energy of the form
\begin{equation}
\label{Visoc}
\hat{V}^{\text{iso}}_{\rm c}(\bar{x})= \hat{\bar{V}}^{\text{iso}}_{\textrm{2-body}}(\bar{x}) +  \bar{V}^{\text{iso}}_{\textrm{many-body}}(\bar{x}), 
\end{equation}
which includes the two-body contribution
\begin{eqnarray}
\label{Visoc2b}
\hat{\bar{V}}^{\text{iso}}_{\textrm{2-body}}(\bar{x})=
\left[
\begin{matrix}
& \bar{V}_{1}^{\rm iso}(\bar{x})    & \bar{K}_{12}^{\text{iso}}(\bar{x}) &\cdots  & \bar{K}_{1f}^{\text{iso}}(\bar{x}) \\
& \bar{K}_{12}^{\text{iso}}(\bar{x})      & \bar{V}_{2}^{\text{iso}}(\bar{x})   &\cdots  & \bar{K}_{2f}^{\text{iso}}(\bar{x}) \\
&\vdots &\vdots &  \ddots &\vdots \\
& \bar{K}_{1f}^{\text{iso}}(\bar{x})   & \bar{K}_{2f}^{\text{iso}}(\bar{x})  &\cdots & \bar{V}_{f}^{\text{iso}}(\bar{x}) \\
\end{matrix}
\right]
\end{eqnarray}
for which the diagonal terms are the centroid potential of mean force for each diabatic surface, 
\begin{eqnarray} 
e^{-\beta \bar{V}_{i}^{\rm iso}(\bar{x})} \! &=& \! \lim_{n\rightarrow\infty} C \int \!\!\!d\textbf{x}\ \delta (\bar{x}-\frac1n \!\sum_{\alpha} x_{\alpha})\\
&&\times \textrm{exp}\left[ -\beta \left(
U_{\text{spr}}(\textbf{x}) + \frac1n \sum_{\alpha=1}^n V_i(x_{\alpha})\right)\right]\nonumber
\end{eqnarray}
for $i=1,\ldots,f$, and the off-diagonal terms are given by
\begin{eqnarray} 
\left(\bar{K}_{ij}^{\text{iso}}(\bar{x})\right)^2  &=&  {\rm acosh}^2 \left[ \, e^{\frac{\beta}{2} \left( \bar{V}_{i}^{\text{iso}}(\bar{x}) +\bar{V}_{j}^{\text{iso}}(\bar{x}) \right)} \, \bar{\mu}_{ij}(\bar{x}) /2   \right] / \beta^2 \nonumber\\ 
&&\quad
 - \left(  \bar{V}_{i}^{\text{iso}}(\bar{x}) -\bar{V}_{j}^{\text{iso}}(\bar{x}) \right)^2 / 4,
\end{eqnarray}
where
\begin{equation} 
 \bar{\mu}_{ij} (\bar{x})
 =   \lim_{n\rightarrow\infty} C \! \int \!\! d\textbf{x} \ \delta (\bar{x}-\frac1n \!\sum_{\alpha} x_{\alpha})\, e^{-\beta U_{\text{spr}}(\textbf{x})} \, \mu_{ij}(\textbf{x}),
\end{equation}
and $\mu_{ij}(\textbf{x})$ is given by Eq.~\ref{muij}.  Also included in the isomorphic potential is the many-body contribution,
\begin{equation}
\bar{V}^{\text{iso}}_{\textrm{many-body}}(\bar{x}) 
=-\frac1\beta\ \textrm{ln} \left[\frac{ \bar{\mu}(\bar{x})}
{ \textrm{tr}_{\textrm{e}}\!\left[  e^{-\beta \hat{\bar{V}}^{\text{iso}}_{\textrm{2-body}}(\bar{x}) } \right]   } \right],
\end{equation}
which vanishes for the case of a two-level system.

\section{The positivity and evaluation of $\mu$}

In the limit of large bead number, $\mu$ can be expressed as a continuous path integral
\begin{eqnarray}
\label{b1}
\lim_{n \rightarrow \infty} \mu(\textbf{x})
&&=\lim_{n \rightarrow \infty} 
{\rm tr_e} \left[ 
\prod_{\alpha=1}^{n} e^{-\beta_n \hat{V}(x^{(\alpha)}) } 
\right] \\
&&=  
{\rm tr_e} \left[ 
\exp_{(\hat{\text{O}})} \left( {-\int_0^\beta \hat{V}(x(\tau)) d \tau}  \right)
\right],\nonumber 
\end{eqnarray}
where 
$\exp_{(\hat{\text{O}})}$ is the time-ordered exponential, which is needed since 
$\hat{V}(x)$ may not commute with itself at different imaginary times along the path, $x(\tau)$. 
Application of the generalized cumulant expansion\cite{kubo1962generalized}  to this  time-ordered exponential yields 
\begin{eqnarray}
\label{b2}
\lim_{n \rightarrow \infty} \mu(\textbf{x}) = \exp  \left( \,\,\, \sum_{j=1}^{\infty} (-1)^j \,\, K_j(\textbf{x}(\tau)) \, \right),
\end{eqnarray}
where  $K_j$ is the $j^{\rm{th}}$-order cumulant 
\begin{eqnarray}
K_j(\textbf{x}(\tau))
=&& \int_0^{\beta}  d\tau_1 \int_0^{\tau_1} d\tau_2 \cdots \int_0^{\tau_{n-1}} d\tau_j\, \nonumber\\
&&\quad
{\rm tr^{(c)}_e} \left[
  \hat{V}(\textbf{x}(\tau_1)) \cdots \hat{V}(\textbf{x}(\tau_j))  
\right],
\end{eqnarray}
and  ${\rm tr^{(c)}_e} \left[\cdot\right]$ is the cumulant partial trace defined in Eq.~2.9 of Ref.~\onlinecite{kubo1962generalized}.
Given that the exponand in Eq.~\ref{b2} is thus a sum of real numbers, it follows that 
$\lim_{n \rightarrow \infty} \mu(\textbf{x})>0$.

In practice, for the $n$-bead discretization of the path integral, 
both $\mu$ and its derivatives $\partial \mu / \partial x^{(\alpha)}$ are evaluated using Bell's algorithm,\cite{bell2005unpublished} which requires only $\mathcal{O}(n)$ operations.  Details of this algorithm are provided elsewhere.\cite{menzeleev2014kinetically, hele2005an}

\section{Time-evolution of the initial phase-space distribution under QCLE dynamics}
\label{appendixQCLE}

\begin{figure}
\includegraphics[width=8.4cm]{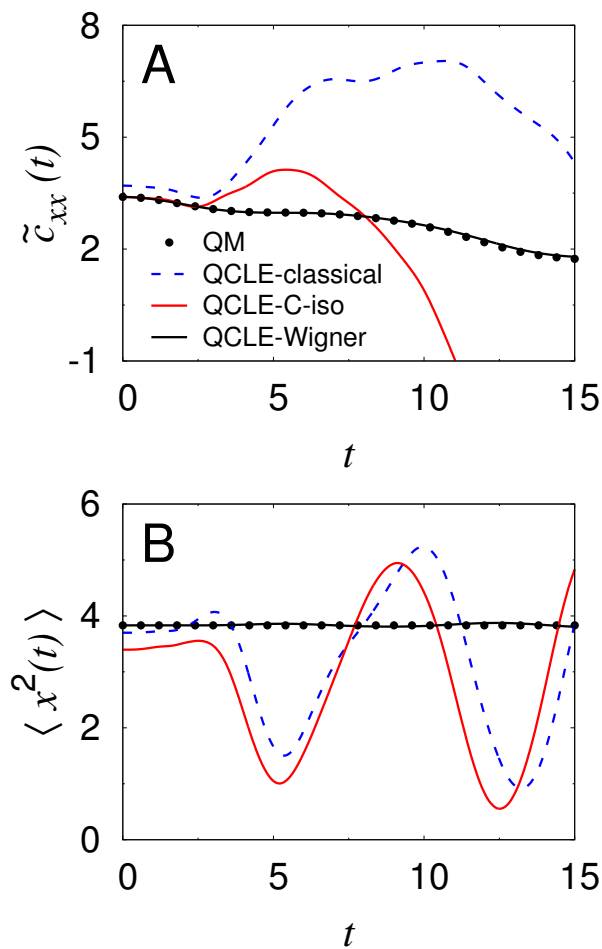}
\caption{\label{fig_4}
{\bf(A)} Kubo-transformed position-autocorrelation functions for two linearly coupled harmonic oscillators, with physical potential energy matrix elements of $V_1(x)=\frac12 (x-x_0)^2$, $V_2(x)=\frac12 (x+x_0)^2$, and $K_{12}(x)=1.25$, where  $x_0=2$,  and $\beta=8$.
{\bf(B)} Time-evolution of the second moment of the phase-space distribution with respect to position, $\langle x^2(t)\rangle$, for the system in panel A.
Results are obtained using exact quantum dynamics (QM; black, dots), QCLE dynamics with  nuclei initialized from the classical phase-space distribution on the physical potential (QCLE-classical; blue, dashed), QCLE dynamics with 
nuclei initialized from the classical phase-space distribution on the isomorphic Hamiltonian (QCLE-C-iso; red, solid), and QCLE dynamics with  nuclei initialized from the multi-surface partial Wigner distribution (QCLE-Wigner; black, solid). 
}
\end{figure}

Here, we examine a source of error for the QCLE dynamics presented in Fig.~\ref{fig_1}E-H of the main text. % {\color{blue}{, similar to the previous comparison between LSC-IVR and Matsubara dynamics [cite]}}.
In particular, we quantify the extent to which the QCLE dynamics preserves the MQC phase-space distribution that arises in the classical limit for the nuclear degrees of freedom (Section~\ref{Sec:Methods:QCLE}).
For a two-level system comprised of linearly coupled one-dimensional harmonic oscillators (see caption), Fig.~\ref{fig_4}A shows results for the Kubo-transformed position-autocorrelation function, and  Fig.~\ref{fig_4}B shows the second moment of the time-evolved initial phase-space distribution with respect to position, $\langle x^2(t)\rangle$.  

The results in Fig.~\ref{fig_4}A are similar to those discussed in Fig.~\ref{fig_1}E-H, with substantial errors emerging for both the QCLE-classical and QCLE-C-iso  at lower temperature and higher coupling; as is necessary for the system studied in this appendix,\cite{mac2002surface} the QCLE dynamics initialized from the multi-level partial Wigner distribution (QCLE-Wigner in Fig.~\ref{fig_4}A) recovers exact quantum mechanics.
As is seen in panel B, the QCLE dynamics exactly preserves the second moment of the initial Wigner phase-space distribution for this system,\cite{mac2002surface, nielsen2000mixed} but it does not preserve the initial MQC phase-space distribution associated with either the physical potential (QCLE-classical) or the isomorphic potential (QCLE-C-iso).  Indeed, the erroneous features in the time correlation functions in panel A coincide with non-conservation of the MQC phase-space distribution in panel B.

Although use of an initial MQC phase-space distribution for the QCLE dynamics emerged (Section~\ref{Sec:Methods:QCLE}) from
our goal of obtaining a classical limit for the nuclear degrees of freedom without double-counting of nuclear quantum effects from the initial distribution, 
it is clear that the MQC phase-space distribution is not conserved by the QCLE dynamics, leading to erroneous time correlations in both the QCLE-classical and QCLE-C-iso results.

\nocite{*}
\bibliography{reference}

\end{document}